\documentclass[a4paper]{PoS}

\usepackage{amsmath}
\usepackage{amssymb}
\usepackage{subcaption}
\usepackage{units}
\usepackage{siunitx}
\usepackage{url}

\title{Results of the EUSO-SPB1 flight}

\ShortTitle{Results of the EUSO-SPB1 flight}

\author{\speaker{J. Eser}$^1$\thanks{jeser@mines.edu} ,
A. Olinto$^2$,
L. Wiencke$^1$\\

$^{1}$Colorado School of Mines\\
$^{2}$University of Chicago
}

\author{for the JEM-EUSO Collaboration \footnote{for collaboration list see PoS(ICRC2019)1177}}

\abstract{The latest and most advanced effort towards a space-based optical cosmic ray detector developed within the Joint Experiment Mission for the Extreme Universe Space Observatory (JEM-EUSO) collaboration was the Extreme Universe Space Observatory on a Super Pressure Balloon (EUSO-SPB1) mission. The EUSO-SPB1 instrument looks for UV light emitted by extensive air showers above the detectors energy threshold of \unit[3]{EeV}.\\
This detector was launched in 2017 out of Wanaka, New Zealand as a mission of opportunity on a NASA SPB. Over 27 hours of data was taken in air shower detection mode during the 12-day flight over the Pacific Ocean.\\
Besides an overview of the instrument and the mission details, we will show the results of the data analysis of the flight. Methods to search for tracks and other interesting signals were developed and applied to the flight data set revealing different types of events. But no obvious track of a cosmic ray candidate was found. This result is in agreement with a detailed simulation study performed after the flight to include the different conditions. Data of the flown IR camera and weather forecast model were used to determine the cloud conditions within the telescopes FoV. The presented results are also discussed in various separate contributions at this conference. The experience gained during this flight is essential for the preparation of the follow-up mission EUSO-SPB2 which is planned to launch in 2022.}

\FullConference{36th International Cosmic Ray Conference -ICRC2019-\\
		July 24th - August 1st, 2019\\
		Madison, WI, U.S.A.}

\begin{document}

\section{Introduction}
Ultrahigh energy cosmic rays (UHECRs) with a measured record energy of \unit[3]{EeV} are the most energetic particles known to exist but their origin and acceleration mechanism remain unknown. Cosmic rays (CRs) at the highest energy can provide a new window to the Universe and add important information to the multi-messenger view of the cosmos.\\
As the flux at these energies is too low for direct measurements, indirect measurement techniques are required using the Earth's atmosphere as a calorimeter. A CR entering the atmosphere deposits its energy by interacting with particles in the atmosphere creating a cascade of secondary particles, a so-called Extensive Air Shower (EAS). Charged particles within the EAS excite nitrogen molecules in the air. The fluorescence light emitted by the de-excitation can be measured and used to characterize the primary CR as successfully proven from ground by Fly's Eye \cite{FlyEye}, HiRes \cite{HiRes}, the Pierre Auger Observatory \cite{PAO} and Telescope Array (TA) \cite{TA}.\\
To overcome the low flux and to observe the whole sky with one detector the approach has to be changed. A UV fluorescence detector in space could be such a new approach and would be not only sensitive to UHECR but also to high-energy photons and cosmogenic tau neutrinos. The Extreme Universe Space Observatory on a Super Pressure Balloon (EUSO-SPB1) \cite{SPB1} was an important step towards the realization of such an instrument.\\
EUSO-SPB1 was the first mission capable of detecting EAS from suborbital space looking down on the atmosphere build by the collaboration. This prototype was launched on April 24th 23:51 UTC 2017 from Wanaka, New Zealand as the payload of a NASA Super Pressure Balloon test flight. The main science goals besides technology development were:
\begin{itemize}
\item First recording of an EAS produced by an UHECR with a fluorescence detector looking down on the atmosphere from suborbital space.
\item UV background estimation at night over the ocean and clouds.
\item Search for fast UV pulse-like signatures from other objects.
\end{itemize}
\section{EUSO-SPB1 Instrument}

\begin{figure}
\centering
\includegraphics[width=.7\textwidth]{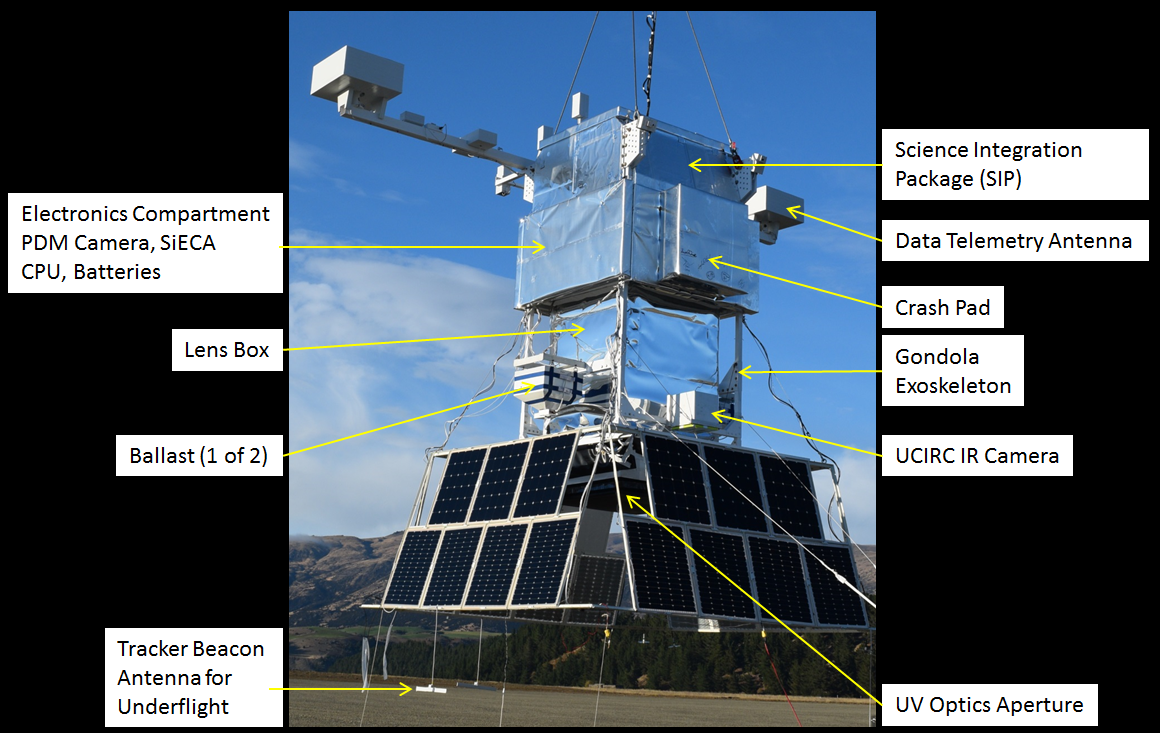}
\caption{The EUSO-SPB1 instrument in flight configuration \cite{EUSOSPB1instrument}.}
\label{fig:EUSOSP1instrument}
\end{figure}

The EUSO-SPB1 instrument (fig. \ref{fig:EUSOSP1instrument}) design was based on an earlier prototype, EUSO-Balloon, flown in 2014 as an overnight flight conducted by the French Space Agency (CNES) \cite{EUSOballoonPeter,EUSOballoonMario}. The balloon was launched from Timmins, Canada, and the flight lasted for around 8 hours. The camera was triggered with a \unit[20]{Hz} frequency, recording UV light background levels \cite{bg-paper} as well as artificial sources (UV flashers and UV laser) from a helicopter \cite{helicopter-paper}.
The detector came down in a small lake and the instrument was recovered intact allowing  reuse the instrument after extensive upgrades for the 2017 long duration balloon flight.
The upgrade includes a new set of Fresnel lenses, a new UV camera (higher quantum efficiency), a trigger to identify CR event candidates, interfaces to the NASA science integration package (SIP) and telemetry and a gondola exoskeleton frame including solar panels to name just the most important ones.\\
A detailed description of the instrument is given in \cite{EUSOSPB1instrument}. The heart of the detector is the 2304 pixel Photo Detector Module (PDM) an ultrafast UV camera, mounted at the focal surface of two \unit[1]{m$^2$} PMMA plastic Fresnel lenses. The PDM counts single photoelectrons. It is built up of nine Elementary Cells (ECs) with four Multi-Anode PhotoMultiplier Tubes(MAPMTs) each. BG-3 UV transmitting optical filters, glued to the front of the MAPMTs, narrow the spectral response of the camera to the wavelength bandwidth between \unit[290]{nm} and \unit[430]{nm}.
If a trigger is issued (details of the trigger algorithm can be found in \cite{EUSOSPB1trigger}) 128 consecutive data frames (40 frames prior to the trigger and 88 after the trigger occurred) from a system buffer get saved to an on-board 1 TB raid array. A single data frame consists of a list of the counted photoelectrons within one Gate Time Unit (\unit[1]{GTU} = \unit[2.5]{$\mu$s}) for each pixel.\\
Various auxiliary instruments were flown alongside the main camera. An IR camera \cite{IRcam} was used to monitor the cloud coverage underneath the balloon, an important aspect of the reconstruction of CR and an LED system provided in-flight testing of the camera response. A 256 channel extension to the main camera, based on Silicon-Photomultipliers (SiPMs), was flown as an R\&D test. This extension, called SiECA \cite{SiECA}, was operated in a stand alone sampling mode.

\section{Field Testing}
To ensure high quality of the scientific data collected during the mission it is obligatory to characterize the instrument prior to the launch. EUSO-SPB1 was characterized in two steps, first by laboratory tests for the single component and then a full-scale end-to-end test in the field \cite{SPBcalibration} at the TA site in Delta, UT, U.S.A. As ballooning always bears the risk of losing the payload the field test data is the only certain data taken in the flight configuration.\\
\begin{figure}
\centering
\begin{minipage}[b]{.47\textwidth}
 \centering
   \includegraphics[width=1.\textwidth]{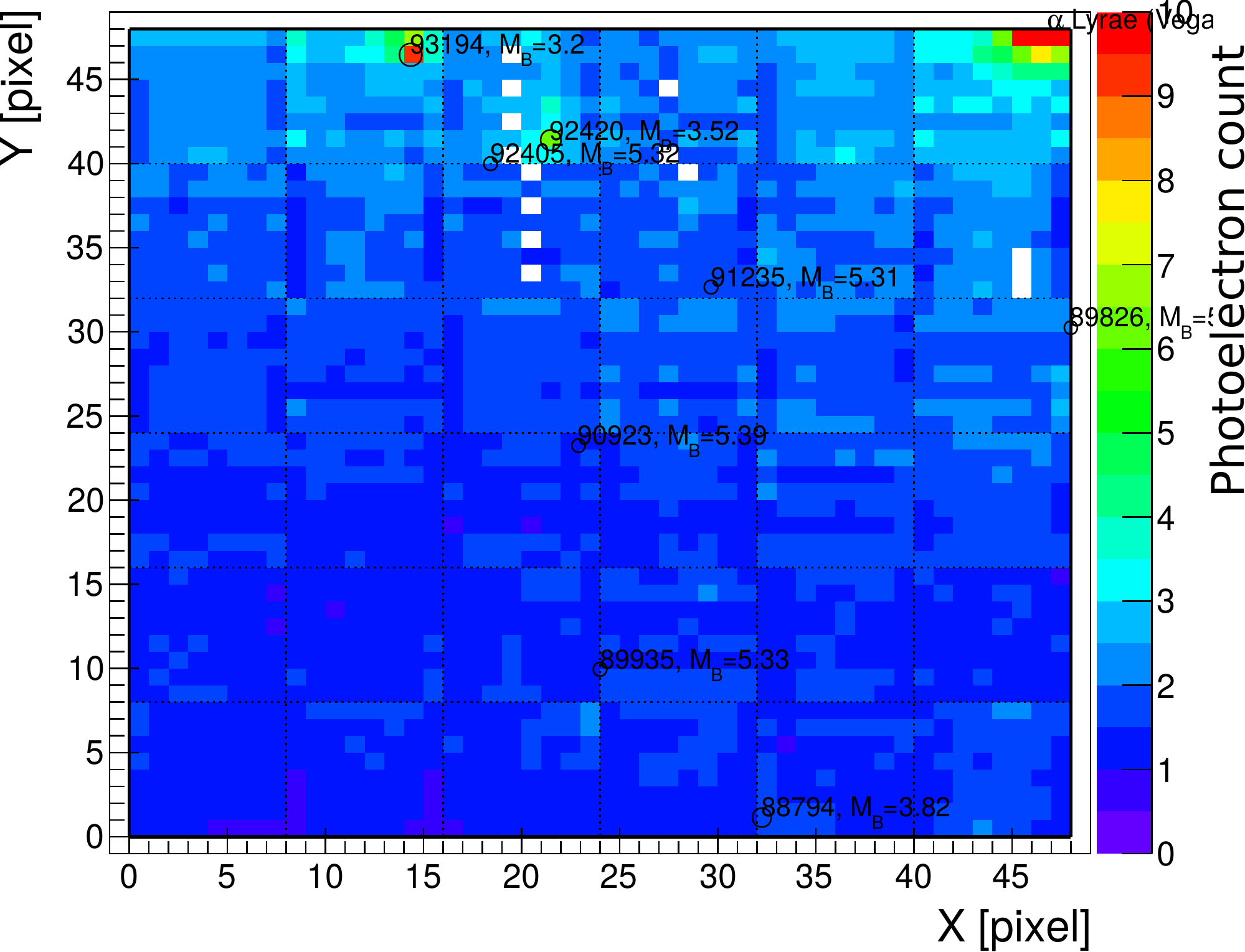}
   \caption{Left: Observed stars used for the FoV measurements \cite{SPBcalibration}.}
   \label{fig:EUSOFoV}
\end{minipage}%
\hfill
\begin{minipage}[b]{.47\textwidth}
 \centering
  \includegraphics[width=\textwidth]{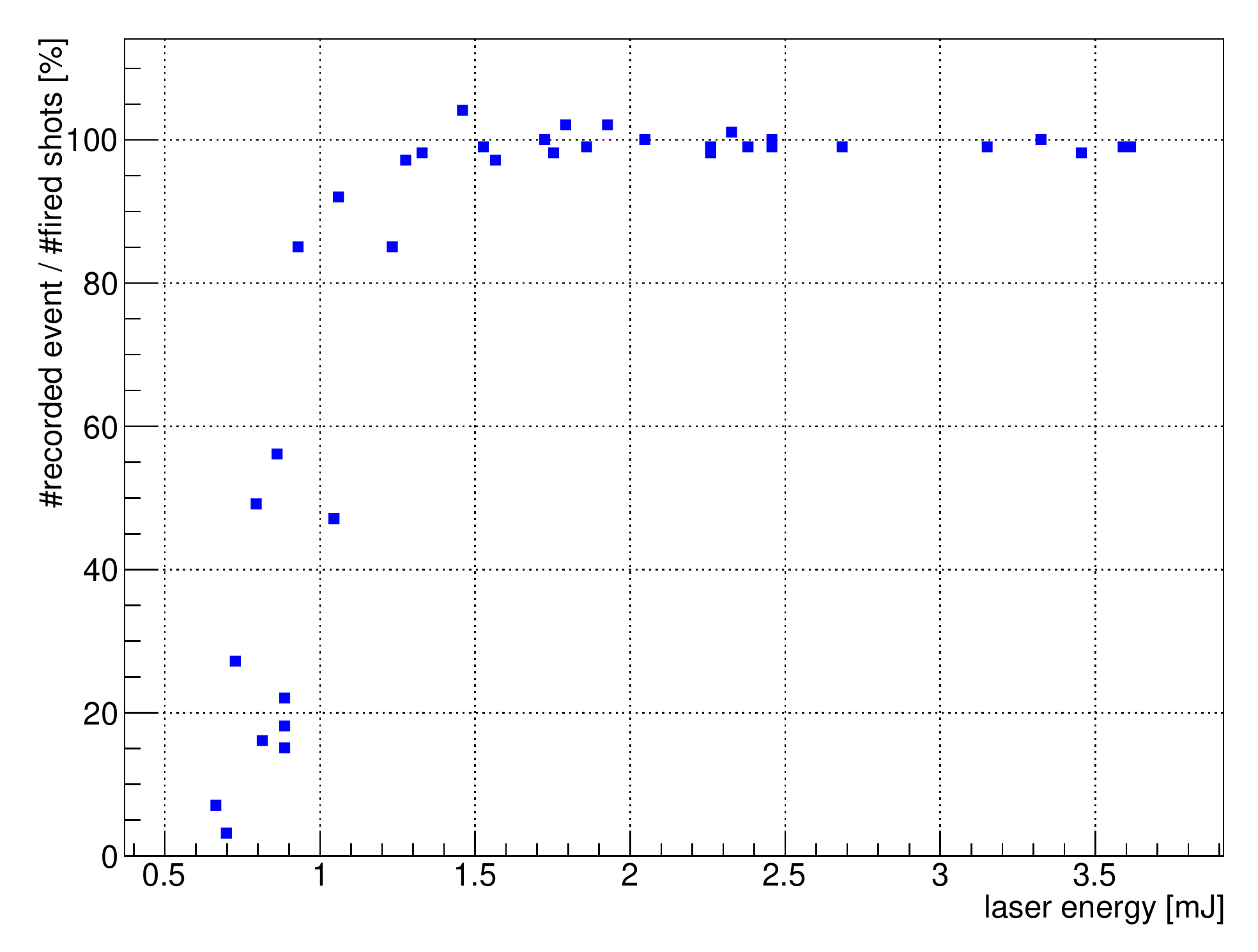}
  \caption{Trigger efficiency for different laser energies \cite{SPBcalibration}.}
  \label{fig:EUSOTrigEff}
\end{minipage}
\end{figure}
A calibrated \unit[365]{nm}-LED was used to perform an absolute calibration of the detector, resulting in an allover efficiency of 0.08 $\pm$ 0.01 pe/photon. This is in agreement with the laboratory measurements. EUSO-SPB1 had a field of view (FoV) of 11.1 $\pm$ \unit[0.2]{$^\circ$}, estimated from the movements of stars within the FoV. This result was confirmed independently by using a steerable laser system (GLS prototype system \cite{PatrikICRC2015}) to produce laser tracks sweeping across the FoV (fig. \ref{fig:EUSOFoV}).\\
Also the trigger efficiency was estimated during this field test. Once again the laser was used to produce EAS like signals in the camera with a well-known energy and direction. We chose a zenith angle of \unit[53]{$^\circ$} and varied the laser energy in a range from \unit[500]{$\mu$J} to \unit[4]{mJ}. The result is shown in fig. \ref{fig:EUSOTrigEff}. The laser energy can be converted to cosmic ray energy establishing the energy detection threshold for EUSO-SPB1 at \unit[3]{EeV} (for a float altitude of \unit[33]{km} and two lenses).

\section{Wanaka Campaign 2017}
\begin{figure}
 \centering
 \includegraphics[width=.9\textwidth]{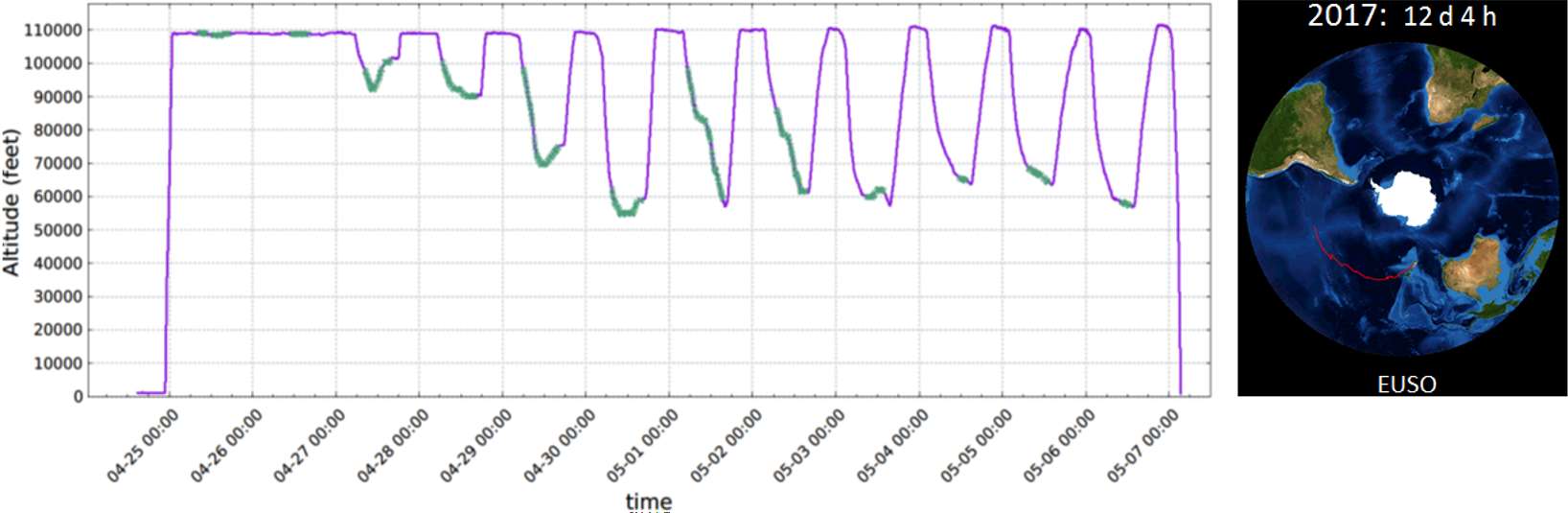}
 \caption{Left: Altitude profile, thick line indicate data taking periods \cite{SPB1}. Right: Flight Trajectory}
 \label{fig:EUSOSPB1SPBflights}
\end{figure}
EUSO-SPB1 was the 3rd Super Pressure Balloon test flight launched from Wanaka, New Zealand. These flights have the goal to reach a  duration of 100 days.
The instrument was declared flight-ready on March 25th 2017 after passing final compatibility and hang tests onsite. The launch occurred on April 24th 23:30 UTC one day before the new moon. The mission was early terminated after the balloon developed a leak and lost the super pressure state (fig. \ref{fig:EUSOSPB1SPBflights}). Only 12 days after the launch, the entire flight train sunk into the ocean about 300 km SE of Easter Island.
During the period of the mission 40 hours of PDM data was recorded. The data was prioritized on board and downloaded via two satellite links. This bandwidth was reduced when one of the two links failed. We were still able to retrieve $\sim$30 hours of PDM data.

\section{Expected Event Rate}
Simulation to estimate the expected event rate conducted prior to the flight had to be repeated afterwards to reflect the actual conditions during the mission including changing float altitudes, trigger adjustment and corrected UV background values. 
\begin{figure}
\centering
\begin{minipage}[b]{.48\textwidth}
 \centering
   \includegraphics[width=1.\textwidth]{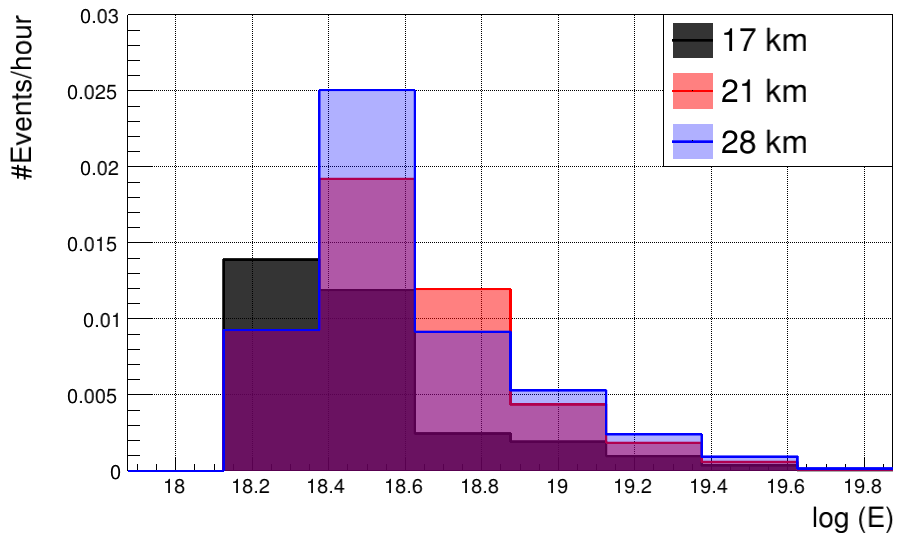}
  \caption{Hourly expected event rate for the EUSO-SPB1 flight for 3 balloon altitudes bin.}
   \label{fig:EUSOSPB1eventrate}
\end{minipage}%
\hfill
\begin{minipage}[b]{.48\textwidth}
 \centering
  \includegraphics[width=1.\textwidth]{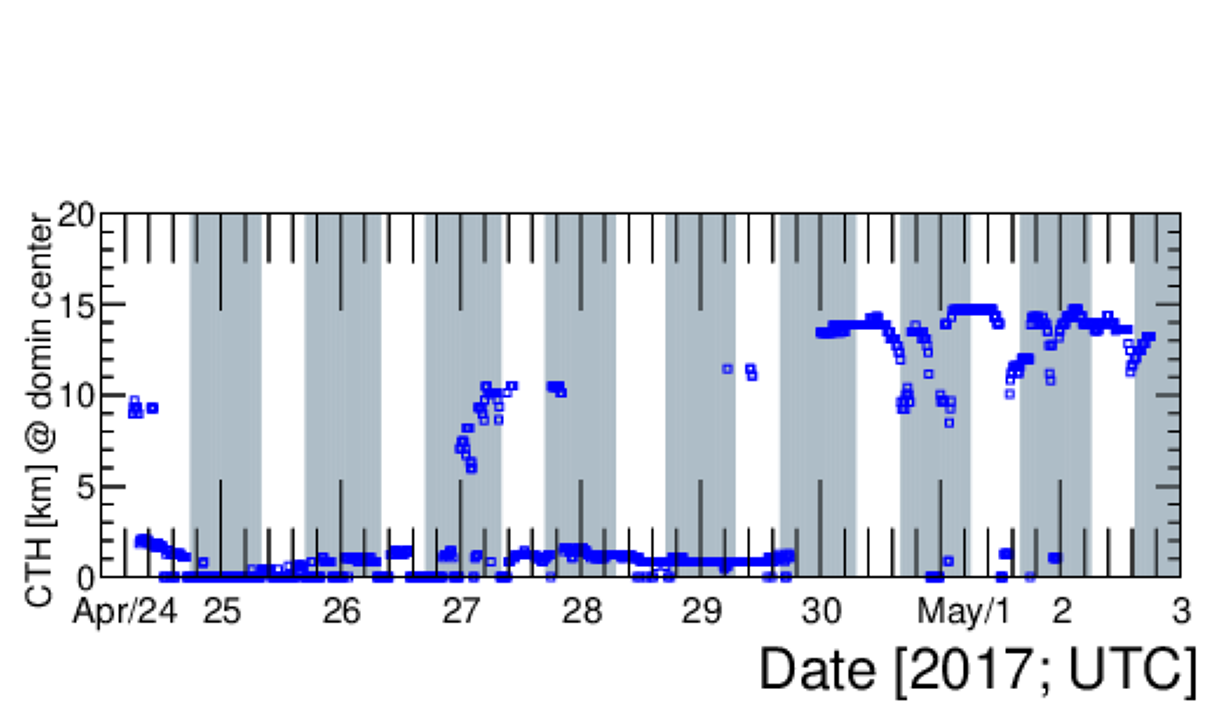}
  \caption{Cloud-top height estimated by a weather forecast model \cite{SilviaICRC2019}.}
   \label{fig:EUSOSPB1cloudCoverage}
\end{minipage}
\end{figure}
Fig. \ref{fig:EUSOSPB1eventrate} displays the event rate per hour for different altitudes based on the cosmic ray flux as presented by the Pierre Auger Observatory \cite{AugerSpectrum}. We simulated an isotropic flux of EAS with protons as primary particles and EPOS-LHC as the hadronic interaction model. The background is based on real flight data. The simulation was tuned by using laser data obtained during the field test to assure an accurate detector representation.\\
To obtain the expected event rate for the entire flight we have to account for the different data taking periods in each altitude bin. From the total data collection time of \unit[1794]{min} (\unit[29.9]{h}) the balloon spend \unit[790]{min} in the \unit[17]{km} bin, \unit[534]{min} in the \unit[21]{km} bin and \unit[470]{min} in the \unit[28]{km} bin. That results in an upper limit on the total number of expected events of 1.23 $\pm$ 0.42 for the flight, assuming a fully working detector over the whole flight and clear atmosphere conditions. As cloud coverage has an impact on the instruments capability at recording EAS, the cloud coverage in the FoV has to be estimated and applied to the event rate calculation. This was done using satellite observation and IR camera pictures when available. Fig. \ref{fig:EUSOSPB1cloudCoverage} shows the cloud top height distribution for the entire flight. From the triggered events at least 60\% fulfill the trigger conditions even when clouds are taken into account. This is true for all energy bins. The cloud corrected event number is 0.74 events.
%At least 60\% of the events are triggered across all energy bins even taken into account the cloud impact. 
Details can be found in these proceedings \cite{KenjiICRC2019} and \cite{SilviaICRC2019}. 

\section{Recorded Data and Cosmic Ray Search}
\begin{figure}
 \centering
 \includegraphics[width=.7\textwidth]{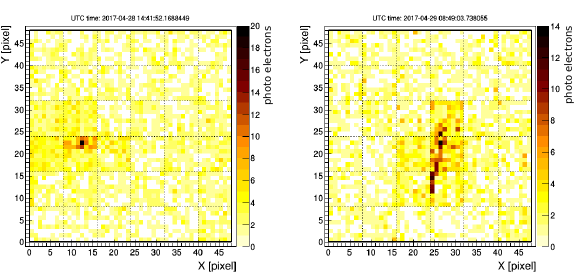}
 \caption{Examples of two types of signal caused by direct cosmic ray hits. Left: circular event. Right: track event}
 \label{fig:directCR}
\end{figure}

\unit[29.9]{h} of PDM data was retrieved from the mission and had to be processed in search of tracks from cosmic ray air showers. The methods range from visual inspection by eye to neural networks.\\ 
%Abraham stuff
%\begin{table}
%\centering
%\begin{tabular}{ c | c | c } 
%Category & Quantity & \% \\ \hline
%Small Blob & 46041 & 51.8 \\
%Edge Effect	& 33802 &	38.1 \\
%PSF Blob & 2995 & 3.4 \\
%Track & 2136 & 2.4 \\
%Hot Pixel & 946 & 1.1 \\
%Big Blob & 710 & 0.8 \\
%Unknown & 2194 & 2.5
%\end{tabular}
%\caption{Event classification of all 88824 triggered events after data reduction based on visual inspection}
%\label{tab:EventClass}
%\end{table}
One analysis approach consisted of observing a subset of the data to understand the kind of events causing a trigger during flight in order to develop an automated method to extract the observed features. 88824 events were extracted and categorized based on the shape of the features \cite{AbrahamICRC2019}. 91\% of the events are small  with a spread of fewer than five pixels, this is smaller than the PSF ruling out that these events are caused by light from outside the telescope. 
To identify air shower tracks a signal length cut of 3 consecutive GTUs was applied, reducing the number of events to 4128. None of these events seem to move in the camera rather are static. The majority of these events start as blobs and decay into single pixels or lines along the edges of the MAPMTs.\\  
%Michal stuff
In another approach, machine learning was used to identify tracks in the recorded data. The goal is to reduce the data set to short and organized lists of events for a manual review. In general, two approaches were investigated. The first approach classifies data based on features extracted by a handwritten algorithm (e.g. Hough Transform) and extremely randomized trees classifiers. The second one utilizes convolutional neural networks trained directly on image data.  A detailed discussion is presented in these proceedings \cite{MichalICRC2019}. Models were trained on simulated air shower events, background noise from the flight, and to increase the accuracy of the method, a manually labeled subset of the triggered flight data. The accuracy on a test data is around 95\%. However, in the worst case, up to 10\% of analyzed flight data subset is classified as air shower candidates.\\ 
Examples of the most common signals found with both approaches are shown in fig. \ref{fig:directCR}. These patterns are produced when a cosmic ray interacts directly with the detector.  
The circular pattern for example could be explained by ionization in the filter glass of the camera or the photo-cathode itself leading to free electrons. More investigation is needed to confirm the process leading to these signals. Both types are lasting only for 1-2 GTUs and moving too fast to be caused by light arriving to the detector from the outside.
Around 60\% of all triggers are related to these types of event making direct cosmic ray hits the main source of false triggers making this data valuable in the development of new trigger algorithms rejecting these type of events.\\
\begin{figure}
 \centering
 \includegraphics[width=0.7\textwidth]{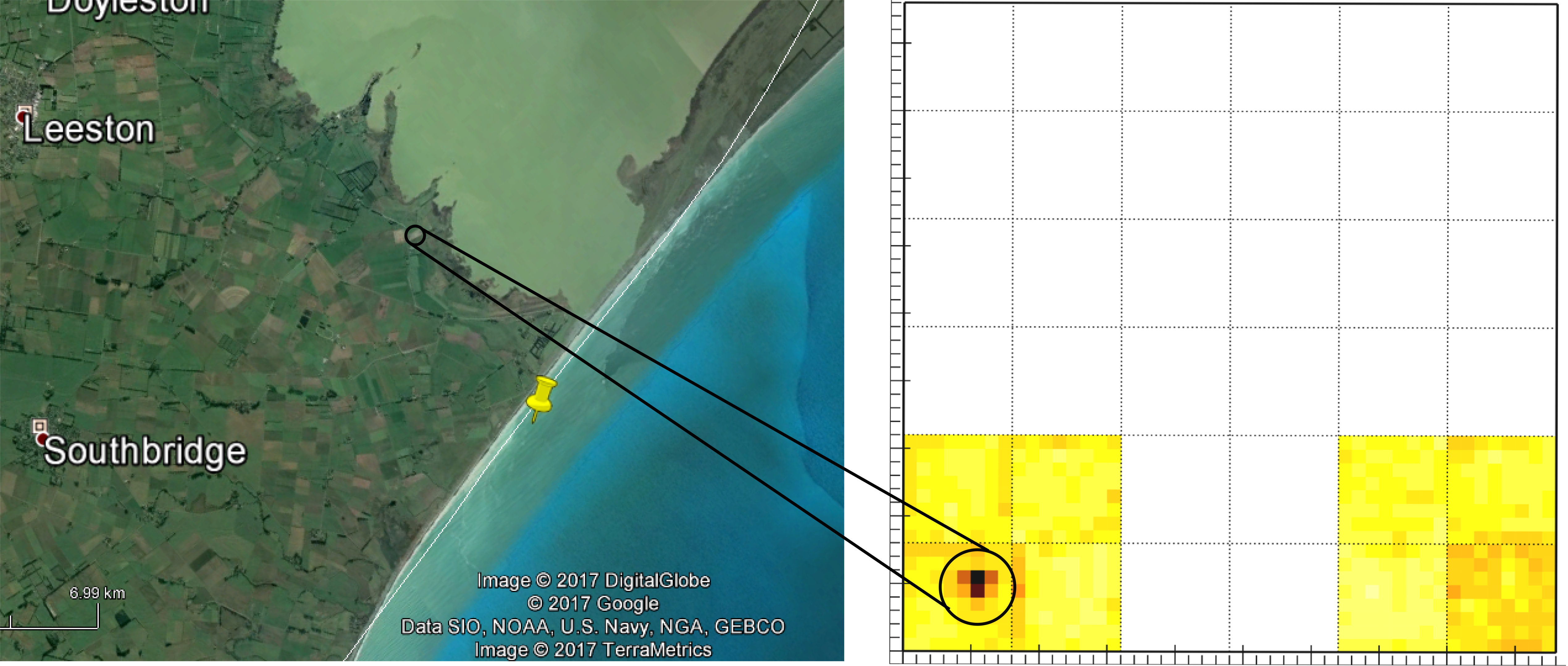}
 \caption{Ground source recorded at 08:32:34 UTC on April 25th, 2017.
 Left: Map of New Zealand's East Coast with black circle indicating the position of the signal. Right: 128 GTU average for the source signal in the detector.}
 \label{fig:EUSOSPB1groundsource}
\end{figure}
In the first night of data taking EUSO-SPB1 recorded a signal (see fig. \ref{fig:EUSOSPB1groundsource}) that was identified to be an artificial light source on ground. The instrument was in commissioning phase at this time with only 2 EC operating. The detection of this light source ensures the maintenance of the optical alignment after the launch as the PSF is the same as measured prior.
%section conclusion
Even though the data was searched with different methods no candidate for a track caused by EAS of a cosmic ray was found. This follows expectation based on new simulations predictions. 

\section{Conclusion and EUSO-SPB2}
Even though the flight was shorter than expected due to an issue with the balloon, EUSO-SPB1 can be viewed as a success. Different types of data show the continuous, stable performance of the camera throughout the flight. New event rate simulation including the different detector and environment condition predict a rate less than one event for the shortened EUSO-SPB1 flight.  Various methods including machine learning, were used to search the data for traces of EAS from UHECR, unfortunately no CR candidate was found agreeing with simulations.\\
The collaboration is in the middle of the preparation for a follow-up mission, EUSO-SPB2 \cite{LawrenceSPB2} with an anticipated launch date of March 2022. This mission expands the science goal to include the measurement of background signals for upward going $\nu_\tau$-events in support of a future space based UHECR observatory like the Probe of Extreme Energy Multi Messenger Astronomy (POEMMA) \cite{POEMMA}. The EUSO-SPB2 payload will consist of two telescopes both utilizing a Schmidt camera design. One is optimized for the detection of EAS from UHECR. The second one is optimized for the Cherenkov light detection of $\nu_\tau$-events by using SiPMs with a very fast readout electronics.

\small{
{\bf Acknowledgment:} This work was partially supported by NASA grants NNX13AH54G, NNX13AH52G, the French Space Agency (CNES), the Italian Space Agency through the ASI INFN agreement n. 2017-8-H.0, the Italian Ministry of Foreign Affairs and International Cooperation, the Basic Science Interdisciplinary Research Projects of RIKEN and JSPS KAKENHI Grant (22340063, 23340081, and 24244042), and the Deutsches Zentrum f\"ur Luft und Raumfahrt, and the 'Helmholtz Alliance for Astroparticle Physics HAP' funded by the Initiative and Networking Fund of the Helmholtz Association (Germany). We acknowledge the NASA Balloon Program Office and the Columbia Scientific Balloon Facility and staff for extensive support, the Telescope Array Collaboration for the use of their facilities in Utah.We thank the Wanaka airport staff and manager Ralph Fegan. We also acknowledge the invaluable contributions of the administrative and technical staff at our home institutions. This research used resources of the National Energy Research Scientific Computing Center (NERSC), a U.S. Department of Energy Office of Science User Facility operated under Contract No. DE-AC02-05CH11231.}


\begin{thebibliography}{99}

\bibitem{FlyEye}
D.J. Bird et al.,
\emph{The Astrophys. J.} {\bf 424}, 491 
(1994).
\bibitem{HiRes}
T. Abu-Zayyad et al., 
\emph{Nucl. Inst. Meth} {\bf A450}, 253 
(2000).
\bibitem{PAO}
J. Abraham et al., [Pierre Auger Collaboration],
\emph{Nucl. Inst. Meth} {\bf A789}, 172 
(2015).
\bibitem{TA}
H. Tokuno et al., [TA Collaboration],
\emph{Nucl. Inst. Meth} {\bf A676}, 54 
(2012).
\bibitem{SPB1}
L. Wiencke et al.,[JEM-EUSO Collaboration],
\emph{EUSO-SPB1 Mission and Science},
in  Proc. of \emph{35th ICRC}, (Busan),
\pos{PoS(ICRC 2017)1097} 
(2017).
\bibitem{EUSOballoonPeter}
P. von Ballmoos et al., [JEM-EUSO Collaboration],
\emph{General overview of EUSO-balloon mission}, 
in  Proc. of \emph{34th ICRC}, (The Hague), 
\pos{PoS(ICRC2015)322}
(2015).  
\bibitem{EUSOballoonMario}
M. Bertaina et al., [JEM-EUSO Collaboration],
\emph{Preliminary results from the EUSO-Balloon flight}
in  Proc. of \emph{34th ICRC}, (The Hague), 
\pos{PoS(ICRC2015)358} 
(2015).
\bibitem{bg-paper}
G. Abdellaoui et al., [JEM-EUSO Collaboration],
\emph{Ultra-violet imaging of the night-time earth by EUSO-Balloon towards space-based ultra-high energy cosmic ray observations},
\emph{Astropart. Phys.} {\bf 111}, 54 
(2019).
\bibitem{helicopter-paper}
G. Abdellaoui et al., [JEM-EUSO Collaboration],
\emph{First observations of speed of light tracks by a fluorescence detector looking down on the atmosphere},
\emph{JINST} {\bf 13}, P05023 
[arXiv:1808.02557]
(2018).
\bibitem{EUSOSPB1instrument}
S. Bacholle et al., [JEM-EUSO Collaboration],
\emph{The EUSO-SPB1 instrument},
in  Proc. of \emph{35th ICRC}, (Busan),
\pos{PoS(ICRC 2017)384} 
(2017).
\bibitem{EUSOSPB1trigger}
J. Bayer et al., [JEM-EUSO Collaboration],
\emph{The trigger logic of EUSO-SPB1 and its expected performance} 
in  Proc. of \emph{35th ICRC}, (Busan),
\pos{PoS(ICRC2017)443}
(2017).
\bibitem{IRcam}
L. Allen et al., [JEM-EUSO Collaboration],
\emph{UCIRC: Infrared Cloud Monitor for EUSO-SPB1} 
in  Proc. of \emph{35th ICRC}, (Busan),
\pos{PoS(ICRC2017)436}
(2017).
\bibitem{SiECA}
W. Painter et al., [JEM-EUSO Collaboration],
\emph{SiECA: Silicon Photomultiplier Prototype for Flight with EUSO-SPB}
in  Proc. of \emph{35th ICRC}, (Busan),
\pos{PoS(ICRC2017)442}
(2017).
\bibitem{SPBcalibration}
J. Eser et al., [JEM-EUSO Collaboration],
\emph{Preflight calibration and testing of EUSO-SPB in the lab and the desert}
in  Proc. of \emph{35th ICRC}, (Busan),
\pos{PoS(ICRC2017)457}
(2017).
\bibitem{PatrikICRC2015}
P. Hunt et al., [JEM-EUSO Collaboration],
\emph{The JEM-EUSO global light system laser station prototype}, 
in  Proc. of \emph{34th ICRC}, (The Hauge),
\pos{PoS(ICRC2015)626}
(2015).
\bibitem{AugerSpectrum}
F. Fenu et al., [Auger Collaboration],
\emph{The cosmic ray energy spectrum measured usingthe Pierre Auger Observatory}, 
in  Proc. of \emph{35th ICRC}, (Busan),
\pos{PoS(ICRC2017)486}
(2017).
\bibitem{KenjiICRC2019}
K. Shinozaki et al., [JEM-EUSO Collaboration],
\emph{An estimation of the exposure of air shower detection by the EUSO-SPB1 mission}, 
in  Proc. of \emph{36th ICRC}, (Madison),
\pos{PoS(ICRC2019)}
(2019).
\bibitem{SilviaICRC2019}
S. Monte et al., [JEM-EUSO Collaboration],
\emph{WRF and radiative methods for Cloud Top Height
retrieval along EUSO-SPB1 trajectory}, 
in  Proc. of \emph{36th ICRC}, (Madison),
\pos{PoS(ICRC2019)}
(2019).
\bibitem{AbrahamICRC2019}
A. Diaz et al., [JEM-EUSO Collaboration],
\emph{EUSO-SPB1: Flight data classification and Air shower search results}, 
in  Proc. of \emph{36th ICRC}, (Madison),
\pos{PoS(ICRC2019)}
(2019).
\bibitem{MichalICRC2019}
M. Vrabel et al., [JEM-EUSO Collaboration],
\emph{Machine Learning
Approach for Air Shower Recognition in the EUSO-SPB1 Experiment Data}, 
in  Proc. of \emph{36th ICRC}, (Madison),
\pos{PoS(ICRC2019)}
(2019).
\bibitem{POEMMA}
A. Olinto et al.,
\emph{POEMMA: Probe Of Extreme Multi-MessengerAstrophysics},
in  Proc. of \emph{35th ICRC}, (Busan),
\pos{PoS(ICRC2017)542}
(2017).
\bibitem{LawrenceSPB2}
L. Wiencke et al., [JEM-EUSO Collaboration],
\emph{The Extreme Universe Space Observatory on a Super-Pressure Balloon II Mission}, 
in  Proc. of \emph{36th ICRC}, (Madison),
\pos{PoS(ICRC2019)}
(2019).
\end{thebibliography}
\end{document}